# BROWNIAN OSCILLATORS DRIVEN BY CORRELATED NOISE IN A MOVING TRAP

Lukáš Glod* – Gabriela Vasziová** – Jana Tóthová** – Vladimír Lisý**

*Department of Mathematics and Physics, The University of Security Management,
Kukučínova 17, 040 01 Košice, Slovakia
**Department of Physics, Faculty of Electrical Engineering and Informatics,
Technical University of Košice, Park Komenského 2, 042 00 Košice, Slovakia

Brownian oscillator, i.e. a micron-sized or smaller particle trapped in a thermally fluctuating environment is studied. The confining harmonic potential can move with a constant velocity. As distinct from the standard Langevin theory, the chaotic force driving the particle is correlated in time. The dynamics of the particle is described by the generalized Langevin equation with the inertial term, a coloured noise force, and a memory integral. We consider two kinds of the memory in the system. The first one corresponds to the exponentially correlated noise in a weakly viscoelastic fluid and in the second case the memory naturally arises within the Navier-Stokes hydrodynamics. Exact analytical solutions are obtained in both the cases using a simple and effective method not applied so far in this kind of problems.

K e y w o r d s: Brownian oscillator, correlated noise, moving harmonic trap

## 1 INTRODUCTION

More than a century has passed since the explanation of the Brownian motion (BM) and creation of its basic theory [1 - 3]. In the works by Einstein, Smoluchowski and Langevin, this phenomenon was mathematically described through stochastic differential equations, in which the influence of the molecules of a surrounding medium on the Brownian particle (BP) was depicted as an additive thermal noise (a fluctuation force) in the equations of motion for the BP. The ever-present thermal fluctuations essentially affect the motion of small (micron-sized or smaller) particles. Although the dynamics of such particles has been studied in an enormous number of works, it still cannot be considered as fully understood [4]. For instance, even in the case of the BM in an unbounded medium the commonly known theory can be applied only for long observation times since it ignores the inertial and accompanying memory effects in the particle motion. With the ongoing trends to miniaturization and interest to systems at small space and time scales, the investigations of these effects become increasingly important. Additional problems arise in studying the systems under spatial confinement. Such situations are realized, e.g., for colloidal particles in optical traps [5]. In several papers the experiments on a BP confined in a moving harmonic well have been theoretically described [5 - 7]. In the case of a BP in a potential well we speak about a Brownian (or noisy) oscillator [3]. For a colloidal particle in a solvent the mentioned inertial effects should necessarily be taken into account at short times when the expected dynamics is ballistic. Also at long times the mean square displacement (MSD) of the particle can exhibit an "anomalous" (different from that in the Einstein theory) time dependence. In the present paper we describe the BM with memory, using the generalized Langevin equation (LE) [1, 2]. The application of the Langevin-Vladimirsky rule [8, 9] allowed us to exactly solve this integro-differential equation for a Brownian oscillator in a harmonic well in two important models. First, we consider the case when the memory kernel in the LE exponentially decreases with the time. According to the second fluctuation-dissipation theorem (FDT) [10], such BPs are driven by coloured, exponentially correlated noise (an Ornstein-Uhlenbeck process). In the limit of zero correlation time of this noise force the obtained solution should agree with the solution of the standard LE. The next case corresponds to the so called hydrodynamic memory in the system, when the memory integral in the LE represents a convolution of the particle acceleration and the memory kernel that naturally arises as a solution of the non-stationary hydrodynamic (Navier-Stokes) equations for incompressible fluids [11]. The history of the hydrodynamic BM begins with the little known work by Vladimirsky and Terletzky [12], who were the first to discover the now famous long-time tails of the particle velocity autocorrelation function (VAF). At the late sixtieth and early seventieth of the last century these tails have been rediscovered in computer experiments, and later they have been confirmed theoretically and experimentally. This discovery doubted the commonly accepted view on the microscopic and macroscopic properties of liquids as being characterized by very different time scales, and extended the range of the applicability of classical hydrodynamics. The hydrodynamic approach has essentially enriched the classical Einstein theory valid only for $t \to \infty$. It has also revealed the limits of its later generalization for arbitrary times (for more details and references see [13, 14]). Such a generalization was made by Langevin who proposed the first stochastic differential equation for the description of the memoryless BM. In the hydrodynamic theory the LE is modified to take into account a possible memory in the particle motion. In the present contribution we obtain the exact solutions to the LE with hydrodynamic memory for a free [12, 15] and confined [16] BP. As distinct from previous works where the inertial or memory effects on the particle motion have been taken into account, in both the considered models the harmonic trap, in which the BP is placed, can move with a constant velocity. It is essential that we use a method of the evaluation of the time correlation functions, such as the VAF and the MSD, which is much simpler than the approaches used in the literature so far and is applicable for linear systems with any other kind of memory.

## 2 OSCILLATOR DRIVEN BY EXPONENTIALLY CORRELATED NOISE

If the random force driving the particles is not the delta-correlated white noise but a coloured noise, the resistance force against the particle motion cannot be arbitrary (in particular, it cannot be the Stokes one as in the traditional Einstein and Langevin theories) but must obey the FDT. Then the equation of motion along the axis $x$ for the BP has a non-Markovian form of the generalized LE [2] that, for a particle of mass $M$ in a harmonic potential $U = kx^2/2$, is

$$M\dot{\upsilon}(t) + M\int_0^t \Gamma(t-t')\upsilon(t')\mathrm{d}t' + M\omega^2 x(t) = f(t), \qquad (1)$$

where the force $f$ has zero mean and its time correlation function at $t > 0$ is $\langle f(t)f(0)\rangle = k_B T\Gamma(t)$. For weakly viscoelastic fluids (for more details see [17]) the memory in the system is described by the kernel $\Gamma(t) = \omega_M \omega_m \exp(-\omega_m t)$. Here, $\omega = (k/M)^{1/2}$ is the oscillator frequency and $\upsilon(t) = \dot{x}(t)$ is the velocity of the BP. Let the correlated random force $f(t)$ arises from the standard LE $m\dot{u}(t) + \gamma u(t) = \eta(t)$ with the white noise $\eta(t)$ and the friction factor $\gamma$ in the Stokes force proportional to the velocity $u(t)$ of the surrounding particles. The characteristic relaxation times of the particles of mass $m$ and the BP of mass $M$, respectively, are $\tau_m = 1/\omega_m = m/\gamma$ and $\tau_M = 1/\omega_M = M/\gamma$. According to the rule first derived in [8], the stochastic LE can be converted to a deterministic equation for the quantity $V(t) = \dot{\xi}(t)$, where $\xi(t)$ is the particle MSD [7], $V(0) = \dot{\xi}(0) = 0$, and the force $f(t)$ is replaced with $2k_B T$ ($k_B$ is the Boltzmann constant and $T$ is the temperature). Using the Laplace transformation $\mathcal{L}$, this equation for $\tilde{V}(s) = \mathcal{L}\{V(t)\}$ reads

$$\tilde{V}(s) = \frac{2k_B T}{M}\frac{s+\omega_m}{s^3 + \omega_m s^2 + (\omega_m \omega_M + \omega^2)s + \omega_m \omega^2}. \qquad (2)$$

The inverse transformation, found after expanding this expression in simple fractions, is

$$V(t) = \frac{2k_B T}{M}\sum_{i=1}^{3} A_i \exp(s_i t), \qquad (3)$$

and the MSD is obtained by simple integration,

$$\xi(t) = \int_0^t V(t')\mathrm{d}t' = \frac{2k_B T}{M}\sum_{i=1}^{3}\frac{A_i}{s_i}\left[\exp(s_i t) - 1\right]. \qquad (4)$$

Here, $s_i$ are the roots of the cubic polynomial in the denominator of (2) and $A_1 = (s_1 + \omega_m)(s_1 - s_2)^{-1}(s_1 - s_3)^{-1}$. The coefficients $A_2$ and $A_3$ have the same form with the cyclic change of the indexes $1\rightarrow 2\rightarrow 3\rightarrow 1$. For these constants the following relations take place

$$\sum_i \frac{A_i}{s_i} = \frac{\omega_m}{s_1 s_2 s_3} = -\frac{1}{\omega^2}, \quad \sum_i A_i = 0, \quad \sum_i A_i s_i = 1, \quad \sum_i A_i s_i^2 = 0, \quad \sum_i A_i s_i^3 = -(\omega_m \omega_M + \omega^2).$$

They can be used in calculations of the asymptotic behaviour of the solution (4). For $\xi(t)$ at $t \rightarrow 0$ we find (the main term being independent on the driving force intensity)

$$\xi(t) \approx \frac{k_B T}{M} t^2 \left(1 - \frac{\omega_m \omega_M + \omega^2}{12} t^2 + \ldots \right). \tag{5}$$

At long times the asymptote can be written in the form (independent on *m*),

$$\xi(t) \approx \frac{2k_B T}{M\omega^2}\left[1 - \exp\left(-\frac{\omega^2 t}{\omega_M}\right)\right]. \tag{6}$$

In the absence of the harmonic force ($\omega \to 0$), we have the expected result $\xi(t \to \infty) \approx 2k_B T t / \gamma$, which follows also from the exact solution (3) and the properties of $A_i$.

Now, let us take into account the possibility that the harmonic well moves with the velocity $\upsilon^*$ along the axis *x* [3]. The position of the BP will be denoted by $x_t = x + x^*$, where *x* obeys the stochastic LE (1) and $x^*$ is the solution of the inhomogeneous deterministic equation

$$\ddot{x}^* + \int_0^t \Gamma(t-t')\dot{x}^*(t')\mathrm{d}t' + \omega^2 x^* = \omega^2 \upsilon^* t. \tag{7}$$

The full solution obeys the GLE (1) with *x* in the last term on the left hand side replaced by $x_t - \upsilon^* t$ and $\upsilon$ changed to $\upsilon_t = \dot{x}_t$. The solution for $x^*(t)$ with the initial conditions $x^*(0) = \dot{x}^*(0) = 0$ can be easily obtained in a similar way as above. Using the Laplace transformation we obtain $x^*(t) = \omega^2 \upsilon^* \sum_i A_i s_i^{-1}\left\{[\exp(s_i t)-1]s_i^{-1} - t\right\}$, with the following limits at short and long times, respectively: $x^*(t \to 0) \approx \omega^2 \upsilon^* t^3 / 6$ and $x^*(t \to \infty) \approx \upsilon^* t$. The full MSD of the particle is calculated as $X(t) = \xi(t) + [x^*(t)]^2$.

Now we show that when the correlation time of the stochastic force *f*(*t*) converges to zero, the obtained solution agrees with the classical solution of the standard Einstein-Langevin theory. When $\tau_m \to 0$ ($\omega_m \to \infty$), (2) reduces to

$$\tilde{V}(s) = \frac{2k_B T}{M} \frac{1}{s_2 - s_1}\left(\frac{1}{s-s_2} - \frac{1}{s-s_1}\right), \quad 2s_{1,2} = -\omega_M \left[1 \mp \left(1 - 4\omega^2 \omega_M^{-2}\right)^{1/2}\right].$$

Then the inverse Laplace transformation gives for the time-dependent diffusion coefficient $D(t) = V(t)/2$

$$D(t) = \frac{k_B T}{M} \frac{1}{s_2 - s_1}\left[\exp(s_2 t) - \exp(s_1 t)\right]. \tag{8}$$

The MSD takes the form

$$\xi(t) = \int_0^t V(t')\mathrm{d}t' = \frac{2k_B T}{M} \frac{1}{s_2 - s_1}\left(\frac{1}{s_1} - \frac{1}{s_2} + \frac{\exp(s_2 t)}{s_2} - \frac{\exp(s_1 t)}{s_1}\right). \tag{9}$$

Using $s_1 s_2 = \omega^2$, the limit at $t \to \infty$ of this expression agrees with (6): $\xi(t \to \infty) = 2k_B T / M\omega^2$.

In the absence of the external force ($\omega \to 0$), the roots are $s_1 = 0$ and $s_2 = -\omega_M$, so that

$$V(t) = \frac{2k_B T}{M\omega_M}\left[1 - \exp(-\omega_M t)\right], \qquad (10)$$

and the MSD has the familiar form [2, 8]

$$\xi(t) = \frac{2k_B T}{\gamma}\left[t + \frac{\exp(-\omega_M t) - 1}{\omega_M}\right], \qquad (11)$$

with the Einstein limit $\xi(t) = 2k_B T t/\gamma$ at $t \to \infty$, and the ballistic behavior $\xi(t) \approx k_B T t^2/M$ at $t \to 0$.

Figures 1 and 2 show the numerical calculations of the VAF $\Phi(t)$ and MSD from our general formula (4) in comparison with the result from the standard theory (9), according to the relation $\Phi(t) = \langle \upsilon(t)\upsilon(0) \rangle = \ddot{\xi}(t)/2$ [4].

The parameters for spherical silica particles in water are taken from the recent experiment [18], where the full transition from the ballistic to diffusive motion has been directly observed for optically trapped BPs. Figure 1 demonstrates the significant differences between the VAFs obtained from the memoryless Langevin theory and that from our model with the memory effects, both for a free particle and a particle in a harmonic potential. In the Einstein theory $\Phi(t) = 0$. Figure 2 shows the MSD calculated for the confined particle. Note however that for small stiffness of the potential (k = 280 pNµm$^{-1}$), the effect of confinement in the experiments [18] is not significant within the used timescale.

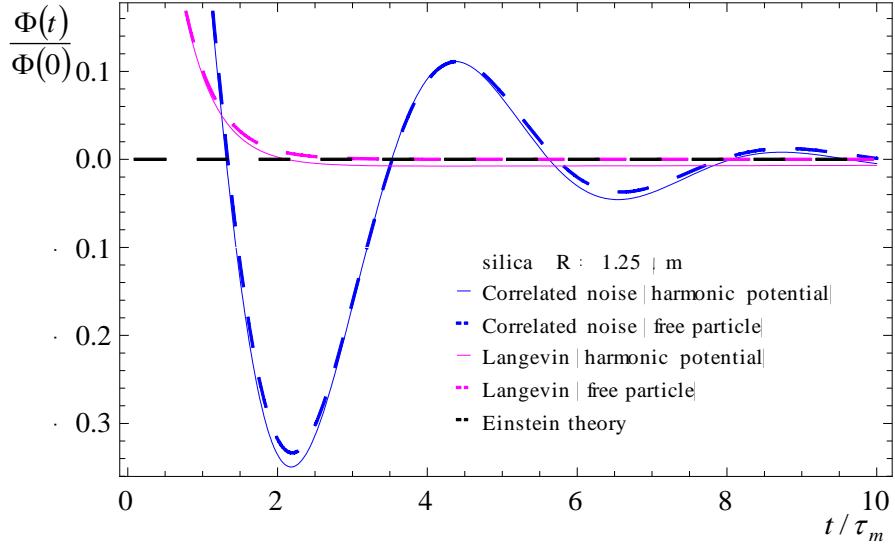

**Fig. 1.** VAF normalized to the $t \to 0$ limit as calculated from (4) and (9) according to the relation $\Phi(t) = \langle \upsilon(t)\upsilon(0) \rangle = \ddot{\xi}(t)/2$ [16]. The parameters are taken from [18] for silica particle in water at room temperature. The value of the harmonic potential parameter $k = M\omega^2$, see Eq.(1), is $k = 280$ pNµm$^{-1}$.

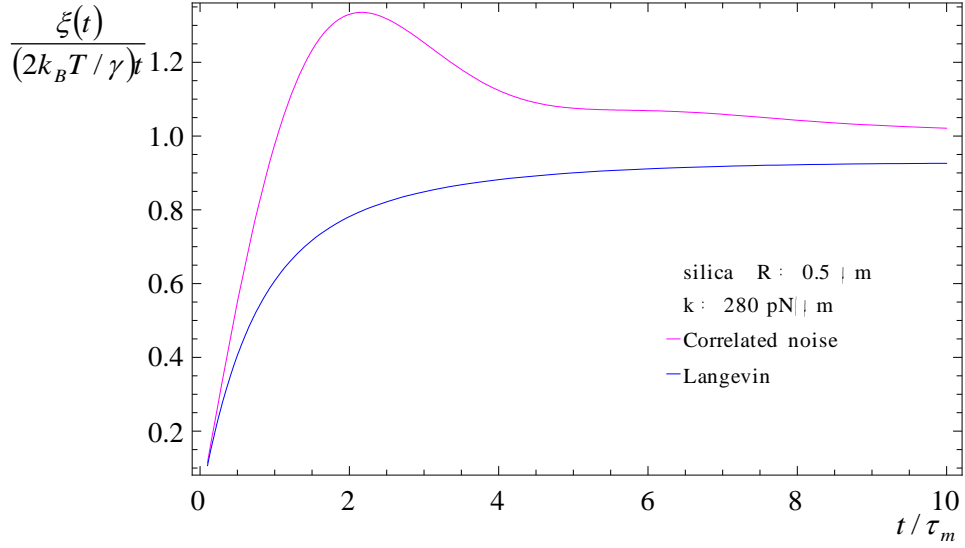

**Fig. 2.** MSD of the Brownian oscillator calculated from (4). The parameters are as in Fig. 1 except the radius of the particle.

## 3 BROWNIAN OSCILLATOR WITH HYDRODYNAMIC MEMORY

The standard LE for the velocity $v(t) = dx/dt$ of a BP has the form

$$m\frac{dv}{dt} = -\gamma v + \sqrt{2D}\xi(t), \tag{12}$$

where the coefficient of friction $\gamma$ for a spherical particle with the radius $R$ and the mass $m$ is the Stokes one, $\gamma = 6\pi R \eta$ ($\eta$ is the dynamic viscosity), and the erratic motion of the particle, resulting from random, uncompensated impacts of the molecules of the surrounding fluid is described by the stochastic (white noise) force ~ $\xi(t)$ with the statistical properties $\langle \xi(t) \rangle = 0$, $\langle \xi(t)\xi(t') \rangle = \delta(t - t')$ and the intensity $D = k_B T \gamma$. The Stokes friction force, which is traditionally used to describe the friction that a particle feels during its motion in a liquid, is in fact valid only for the steady motion of the particle (at long times), and for non-stationary motion should be replaced by the expression [12-15]

$$F(t) = -\gamma\left\{v(t) + \frac{\rho R^2}{9\eta}\frac{dv}{dt} + \sqrt{\frac{\rho R^2}{\pi\eta}}\int_{-\infty}^{t}\frac{dv}{dt'}\frac{dt'}{\sqrt{t-t'}}\right\}, \tag{13}$$

where $\rho$ is the density of the solvent. Equation (13) holds for all times $t \gg R/c$ ($c$ is the sound velocity), i.e., except the very short times when the solvent compressibility must be taken into account. This expression has been derived by Boussinesq in 1885 [19], used in the mentioned work [12], and later brought to wider attention by Hinch [15]. It is seen from (13) that for fluids with the density comparable to the density of the BP (which is the usual case of freely buoyant particles), the terms additional to the Stokes one cannot be neglected since in the equation of motion for the particle they are of the same order as the inertial term. Here we will consider a more complicated problem of the movement of the BP, when the particle is subjected to an external harmonic potential.

The solution can be obtained very easily as follows. Similarly as in the preceding section, using the work [8], instead of (12) with the friction force (13) we can solve the deterministic "equation of motion" for the quantity $V(t) = dX(t)/dt$, where the particle MSD is now denoted as $X(t)$,

$$\dot{V}(t) + \frac{1}{\tau}\sqrt{\frac{\tau_R}{\pi}}\int_0^t \frac{\dot{V}(t')}{\sqrt{t-t'}}dt' + \frac{1}{\tau}V(t) + \omega_0^2\int_0^t V(t')dt' = 2\Phi_0, \qquad \Phi_0 = \frac{k_B T}{M}. \tag{14}$$

Here $M = m + m_s/2$ ($m_s$ is the mass of the solvent displaced by the particle) and $\omega_0^2 = k/M$ ($k$ is is the stiffness of the trapping potential). The characteristic times in this equation are $\tau = M/\gamma$ (the relaxation time of the BP) and $\tau_R = \rho R^2/\eta$ (the vorticity time). The confinement introduces another timescale $\tau_k = \gamma/k$. The constant "force" $2k_B T$ at the right begins to act on the particle at the time $t = 0$; up to this moment the particle is at rest together with the liquid [8, 12]. The problem has to be solved with the evident initial conditions $V(0) = X(0) = 0$. It is also seen from (14) that $\dot{V}(0) = 2\Phi_0$. Taking the Laplace transformation $\mathcal{L}$ of (14), we obtain for $\tilde{V}(s) = \mathcal{L}\{V(t)\}$

$$\tilde{V}(s) = 2\Phi_0 s^{-1}\left(s + \sqrt{\tau_R}\tau^{-1}s^{1/2} + \tau^{-1} + \omega_0^2 s^{-1}\right)^{-1}. \tag{15}$$

Its inversion gives the solution

$$V(t) = 2\Phi_0 \sum_{i=1}^{4} b_i z_i \exp(z_i^2 t)\,\text{erfc}(-z_i\sqrt{t}), \tag{16}$$

where $z_i$ are the roots of the quartic equation $z^4 + \tau_R^{1/2}\tau^{-1}z^3 + \tau^{-1}z^2 + \omega_0^2 = 0$ and the coefficients $b_i$ can be determined decomposing the right hand side of (15) in simple fractions,

$$\tilde{V}(s) = 2\Phi_0 \sum_{i=1}^{4} \frac{b_i}{\sqrt{s} - z_i}, \tag{15a}$$

$$b_1 = \left[z_1^3 - z_1^2(z_2 + z_3 + z_4) + z_1(z_2 z_3 + z_2 z_4 + z_3 z_4) - z_2 z_3 z_4\right]^{-1},$$

$$b_2 = (z_2 - z_1)^{-1}\left[z_2^2 - z_2(z_3 + z_4) + z_3 z_4\right]^{-1},$$

$$b_3 = (z_1 - z_3)^{-1}(z_2 - z_3)^{-1}(z_3 - z_4)^{-1},$$

$$b_4 = (z_1 - z_4)^{-1}(z_2 - z_4)^{-1}(z_4 - z_3)^{-1}.$$

These constants obey the following relations

$$\sum_{i=1}^{4} b_i = 0, \quad \sum_{i=1}^{4} \frac{b_i}{z_i} = -\frac{1}{\omega_0^2}, \quad \sum_{i=1}^{4} \frac{b_i}{z_i^2} = 0, \quad \sum_{i=1}^{4} \frac{b_i}{z_i^4} = \frac{\gamma M}{2k^2}\sqrt{\frac{\tau_R}{\pi}} = \frac{\tau_k}{2\omega_0^2}\sqrt{\frac{\tau_R}{\pi}} = \frac{\tau \tau_k^2}{2}\sqrt{\frac{\tau_R}{\pi}},$$

which can be used in calculations of the asymptotic behaviour of the solution (17), (18), (20) and (25).

Equation (15a) holds for different roots $z_i$. If some of the roots $z_i$ are equal, the coefficients $b_i$ can be obtained by taking the appropriate limits in the above expressions or, alternatively, again by analyzing (15) decomposed in simple fractions. For example, if $z_1 = z_2 \neq z_3 \neq z_4$ and $z_4 \neq z_1$ instead of (15a) we have

$$\tilde{V}(s) = 2\Phi_0 \left[ \frac{b'_1}{\sqrt{s} - z_1} + \frac{b'_2}{\left(\sqrt{s} - z_1\right)^2} + \frac{b'_3}{\sqrt{s} - z_3} + \frac{b'_4}{\sqrt{s} - z_4} \right], \tag{15b}$$

and the coefficients $b'_i$ are obtained by comparing (15b) and (15).

The VAF $\Phi(t) = \langle \dot{x}(t)\dot{x}(0) \rangle = \ddot{X}(t)/2 = \dot{V}(t)/2$ is expressed by a similar equation, if one divides $V(t)$ by 2 and replaces $b_i z_i$ with $b_i z_i^3$. For $\omega_0^2 \to 0$ this expression exactly corresponds to the solutions found in Refs. [12, 15] and contains the long-time tail discovered already in the computer experiments [20, 21]. In our more general case it follows from (16) for the VAF at $t \to \infty$ that

$$\Phi(t) = \Phi_0 \sum_{i=1}^{4} b_i z_i^3 \exp(z_i^2 t) \operatorname{erfc}(-z_i \sqrt{t})$$

$$= -\frac{\Phi_0}{2\sqrt{\pi}} \frac{1}{t^{3/2}} \sum_{i=1}^{4} b_i \sum_{m=1}^{\infty} (-1)^m \frac{(2m-1)!!}{(2z_i^2 t)^{m-1}} \approx \frac{15\Phi_0}{8\sqrt{\pi}} \frac{1}{t^{7/2}} \sum_{i=1}^{4} \frac{b_i}{z_i^4}, \quad \Phi(0) = \Phi_0 \tag{17}$$

the longest-lived tail is $\sim t^{-7/2}$. Finally, the MSD of the BP is found integrating the function $V(t)$ from 0 to $t$,

$$X(t) = 2\Phi_0 \sum_{i=1}^{4} \frac{b_i}{z_i} \left[ \exp(z_i^2 t) \operatorname{erfc}(-z_i \sqrt{t}) - 1 \right], \tag{18}$$

and asymptotic expansion of this equation for $t \to \infty$ is

$$X(t) = -2\Phi_0 \sum_{i=1}^{4} \frac{b_i}{z_i} \left[ 1 + \frac{1}{z_i \sqrt{\pi t}} \sum_{m=1}^{\infty} (-1)^m \frac{(2m-1)!!}{(2z_i^2 t)^m} \right] \approx \Phi_0 \left[ \frac{2}{\omega_0^2} + \frac{1}{\sqrt{\pi}} \frac{1}{t^{3/2}} \sum_{i=1}^{4} \frac{b_i}{z_i^4} \right]. \tag{19}$$

Equation (18) agrees very well with the experiments on free [14] and confined [18] BPs. The corresponding numerical calculations for resin and polystyrene particles in water [18] are shown in Fig. 3 for the VAF and in Fig. 4 for the MSD. The figures demonstrate the significant differences between the standard Langevin and the presented hydrodynamic theory.

The positional autocorrelation function (PAF) of the trapped sphere $\langle x(t)x(0) \rangle$, measured in [22], can be also computed directly. The PAF is derived from (18) via the relation $\langle x(t)x(0) \rangle = k_B T / k - X(t)/2$ as

$$\langle x(t)x(0)\rangle = -\Phi_0 \sum_{i=1}^{4} \frac{b_i}{z_i} \exp(z_i^2 t)\,\mathrm{erfc}(-z_i\sqrt{t}). \tag{20}$$

From here for we have for the PAF at $t \to \infty$

$$\langle x(t)x(0)\rangle = \frac{\Phi_0}{\sqrt{\pi t}}\sum_{i=1}^{4}\frac{b_i}{z_i^2}\sum_{m=1}^{\infty}(-1)^m\frac{(2m-1)!!}{(2z_i^2 t)^m} \approx -\frac{\Phi_0}{2\sqrt{\pi}}\frac{1}{t^{3/2}}\sum_{i=1}^{4}\frac{b_i}{z_i^4}. \tag{21}$$

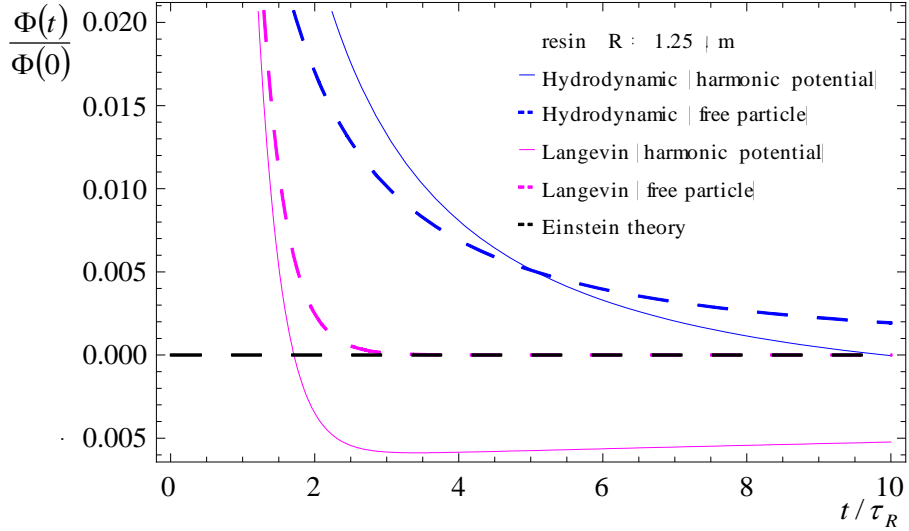

**Fig. 3.** Normalized VAF for Brownian oscillator with hydrodynamic memory calculated from (18) as $\Phi(t)=\langle \upsilon(t)\upsilon(0)\rangle = \ddot{X}(t)/2$ [16], the parameters are taken from Ref [18].

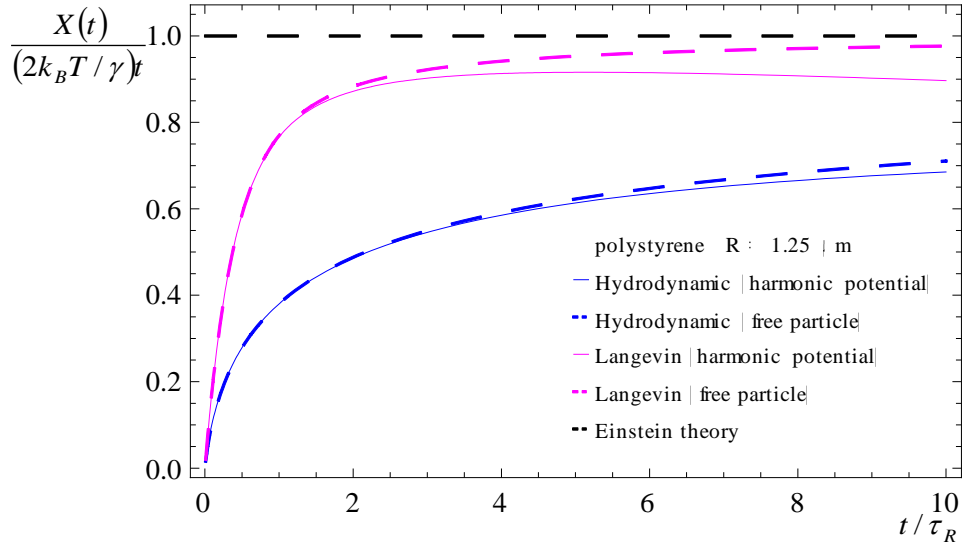

**Fig. 4.** Normalized MSD for Brownian oscillator with hydrodynamic memory calculated from (18). The parameters are from Ref [18].

Finally, let us take into account the possibility that the harmonic well moves with the velocity $v^*$ along the axis $x$. The position of the BP will be denoted by $x_t = x + x^*$, where $x$ obeys the stochastic LE (12) with the Boussinesq force (13), and $x^*$ is the solution of the inhomogeneous deterministic equation

$$\ddot{x}^*(t) + \frac{1}{\tau}\sqrt{\frac{\tau_R}{\pi}} \int_0^t \frac{\ddot{x}^*(t')}{\sqrt{t-t'}} dt' + \frac{1}{\tau}\dot{x}^*(t) + \omega_0^2 x^*(t) = \omega_0^2 v^* t. \qquad (22)$$

The solution for $x^*(t)$ has to be searched for with the initial conditions $x^*(0) = \dot{x}^*(0) = 0$. Taking the Laplace transformation $\mathcal{L}$ of (22), we obtain for $\tilde{x}^*(s) = \mathcal{L}\{x^*(t)\}$

$$\tilde{x}^*(s) = \omega_0^2 v^* s^{-2} \left(s^2 + \sqrt{\tau_R}\tau^{-1}s^{3/2} + \tau^{-1}s + \omega_0^2\right)^{-1}. \qquad (23)$$

The inverse transform is found expanding the therm $(\ldots)^{-1}$ in simple fractions,

$$\tilde{x}^*(s) = \frac{\omega_0^2 v^*}{s^2} \sum_{i=1}^4 \frac{b_i}{\sqrt{s}-z_i}, \qquad (24)$$

which yields

$$x^*(t) = v^* \left\{ t + \omega_0^2 \sum_{i=1}^4 \frac{b_i}{z_i^3} \left[\exp(z_i^2 t)\operatorname{erfc}(-z_i\sqrt{t}) - 1\right] \right\} \qquad (25)$$

with the long-time limit,

$$x^*(t) = v^* \left\{ t - \omega_0^2 \sum_{i=1}^4 \frac{b_i}{z_i^3} \left[1 + \frac{1}{z_i\sqrt{\pi t}} \sum_{m=0}^\infty (-1)^m \frac{(2m-1)!!}{(2z_i^2 t)^m}\right] \right\} \approx v^* \left\{ t - \omega_0^2 \sum_{i=1}^4 \frac{b_i}{z_i^3} \right\}. \qquad (26)$$

The full MSD is

$$X_t(t) = \left\langle \left[x(t) - x(0) + x^*(t) - x^*(0)\right]^2 \right\rangle = X(t) + x^{*2}(t). \qquad (27)$$

## 4 CONCLUSION

The theory of the Brownian motion is still being developed and along with the remarkable improvements of experimental possibilities it finds an increasing number of applications, especially in the science and technology of micro- and nano-systems systems. It has been found that in many situations [1, 2, 14, 18] the standard Langevin equation should be generalized by taking into account the memory effects on the dynamics or, equivalently, the effects of finite correlation time of the noise driving the particles. In our work, a specific problem of the motion of a Brownian particle under the influence of an exponentially correlated stochastic force has been solved within the classical consideration. As distinct from the usual approaches, the inertial effects have been taken into account. We have examined the case of a free particle as well as the motion of a particle in a harmonic trap (a stochastic oscillator with memory). From the generalized Langevin equation the exact velocity autocorrelation function, the time-dependent diffusion coefficient, and the mean square displacement $\xi(t)$ of the particle have been calculated and analyzed in detail. At short times $\xi(t) \sim t^2$ describes the ballistic motion. At $t \to \infty$ it converges to a constant strongly depending on the oscillator frequency $\omega$ and agrees with the Einstein diffusion limit when $\omega \to 0$. The full mean square displacement for a trapped particle corresponds to the experiments on colloids [6]. Our results can be also used to describe the charge fluctuations in nanoscale electric circuits in contact with the thermal bath, for which essentially the same equations have been derived from the first principles [23, 24]. We have also considered a similar problem of the dynamics of a Brownian particle moving in a liquid, when the standard Langevin equation does not represent a good model. In this case the memory follows naturally from the non-stationary Navier-Stokes hydrodynamics of incompressible fluids. We have solved the problem of the hydrodynamic Brownian motion of a particle in an external harmonic potential. The velocity autocorrelation function and the mean square displacement have been obtained in a more effective way than within the approaches used in the literature so far. We have generalized the results known for free and confined particles [16] to the presence of a moving potential well, in correspondence with a number of recent experiments on particles in optical traps.

## Acknowledgement

This work was supported by the Agency for the Structural Funds of the EU within the project NFP 26220120033, and by the grant VEGA 1/0370/12.